\documentclass[conference]{IEEEtran}

\usepackage{amssymb}

\usepackage{cite}

\usepackage{graphicx}
\usepackage{subfigure}

\usepackage{amsmath}

\usepackage{algorithm}
\usepackage{algorithmic}
\usepackage{multirow}
\renewcommand{\algorithmicrequire}{\textbf{Input:}} 

\usepackage{float}

\begin{document}
%
\title{Optimal Scheduling for Incentive WiFi Offloading under Energy Constraint}

\author{\IEEEauthorblockN{Juntao Gao, Minoru Ito}
\IEEEauthorblockA{Graduate School of Information Science\\
Nara Institute of Science and Technology, Nara, Japan\\
Email: \{jtgao, ito\}@is.naist.jp}
\and
\IEEEauthorblockN{Norio Shiratori}
\IEEEauthorblockA{Research Institute of Electrical Communication\\
Tohoku University, Sendai, Japan}}

\maketitle

\begin{abstract}
WiFi offloading, where mobile device users (e.g., smart phone users) transmit packets through WiFi networks rather than cellular networks, is a promising solution to alleviating the heavy traffic burden of cellular networks due to data explosion. 
However, since WiFi networks are intermittently available, a mobile device user in WiFi offloading usually needs to wait for WiFi connection and thus experiences longer delay of packet transmission.
To motivate users to participate in WiFi offloading, cellular network operators give incentives (rewards like coupons, e-coins) to users who wait for WiFi connection and transmit packets through WiFi networks.

In this paper, we aim at maximizing users' rewards while meeting constraints on queue stability and energy consumption. 
However, we face scheduling challenges from random packet arrivals, intermittent WiFi connection and time varying wireless link states.
To address these challenges, we first formulate the problem as a stochastic optimization problem.
We then propose an optimal scheduling policy, named Optimal scheduling Policy under Energy Constraint (OPEC), which makes online decisions as to when to delay packet transmission to wait for WiFi connection and which wireless link (WiFi link or cellular link) to transmit packets on. 
OPEC automatically adapts to random packet arrivals and time varying wireless link states, not requiring a priori knowledge of packet arrival and wireless link probabilities.
As verified by simulations, OPEC scheduling policy can achieve the maximum rewards while keeping queue stable and meeting energy consumption constraint.

\end{abstract}

\IEEEpeerreviewmaketitle

\section{Introduction}
 
With the explosion of smart devices (like smart phones, tablets) in daily life, mountainous data (like high resolution photos and videos) have been generated every day. Global mobile data traffic has already reached $3.7$ exabytes per month at the end of 2015 and will be 30.6 exabytes per month by 2020, i.e., eightfold increase, according to the report of Cisco \cite{Cisco_report}. 
As a result, cellular networks are overloaded, which degrades users' quality of experience (we already experienced long delay, low throughput during peak hours in crowded downtown areas \cite{Passarella_CST14}). 
To improve users' quality of experience, cellular network operators struggle to increase network connection speeds. However, deploying more cellular infrastructures (like base stations) is very expensive and cannot bring reasonable benefits. Recently, offloading mobile data traffic from cellular networks to WiFi networks of low cost and high connection speed (known as WiFi offloading) has been proposed as a promising solution to alleviating the heavy traffic burden of cellular networks \cite{Lee_TON13,Passarella_CST14}.

\begin{figure}[!t]
\centering
\includegraphics[width=3in]{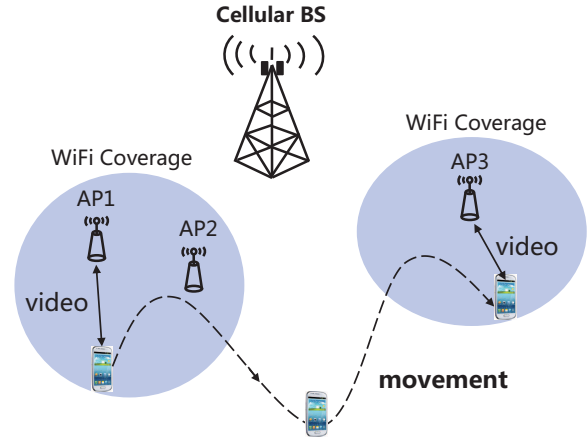}
\caption{Illustration of participants in WiFi offloading.}
\label{fig:WiFioffload}
\end{figure}

Participants in WiFi offloading include cellular network operators (cellular Base Station (BS)), WiFi Access Point (AP) owners and mobile device users, as shown in Figure \ref{fig:WiFioffload}. In such a network, mobile device users leverage loose delay requirement of delay-tolerant applications (e.g., uploading photos and videos to iCloud/youtube, synchronizing dropbox) to postpone packet transmission for utilizing intermittent WiFi connection. While cellular network operators can benefit from WiFi offloading, both WiFi AP owners and users may suffer from bad quality of experience: WiFi AP owners need to sacrifice WiFi bandwidth and energy to offer offloading services and users need to delay packet transmission to wait for intermittent WiFi connection. To motivate WiFi AP owners to provide WiFi offloading services and users to utilize WiFi offloading services, incentive frameworks have been proposed recently (see Section \ref{section:rw} for related work), where cellular network operators give rewards to WiFi AP owners and users to compensate for their loss. However, these frameworks focus on maximizing the utility of network operators (related to operators' satisfaction, e.g., minimizing operators' incentive cost) rather than mobile device users. Further, they do not take into account limited battery power of mobile devices.

In this paper, we aim at maximizing users' rewards in incentive WiFi offloading under constraint on energy consumption of mobile devices. 
However, to achieve this goal, we face scheduling challenges from random packet arrivals, intermittent WiFi connection and time varying wireless link states.
To address these challenges, we first formulate the problem as a stochastic optimization problem with constraint on average energy consumption and queue stability (defined later). 
We then propose an optimal scheduling policy, named Optimal scheduling Policy under Energy Constraint (OPEC), which makes online decisions as to when to delay packet transmission to wait for WiFi connection and which wireless link (WiFi link or cellular link) to transmit packets on. 
OPEC automatically adapts to random packet arrivals and time varying wireless link states, not requiring a priori knowledge of packet arrival and wireless link probabilities.
As verified by simulations, OPEC scheduling policy can achieve the maximum rewards while keeping queue stable and meeting energy consumption constraint.

The rest of this paper is organized as follows. In Section \ref{section:rw}, we review related work on incentive WiFi offloading and scheduling policies under energy constraint. We introduce network models and assumptions for WiFi offloading system in Section \ref{section:sm}. We give a formulation of the optimal scheduling problem in Section \ref{section:pf} and present a detailed description of OPEC scheduling policy in Section \ref{section:OPEC}. Simulation results to verify OPEC are presented in Section \ref{section:sv} and the whole paper is concluded in Section \ref{section:conclusion}.



\section{Related Work} \label{section:rw}

To motivate users to delay their packet transmissions for offloading cellular traffic to complementary networks (DTNs and WiFi hotspots), Zhuo et al. \cite{Gao_TMC14} proposed an incentive framework named "Win-Coupon," in which cellular network operators provide users incentives (service discount coupon). Their objective is to minimize the incentive cost of cellular network operators. 

A similar work \cite{Cai_ICC15}, which also motivates users to offload cellular traffic to WiFi networks, proposed an incentive mechanism to allow network operators to optimally award users, but differs from \cite{Gao_TMC14} in how network operators negotiate an offloading contract with users.

Instead of motivating users, Kang et al. \cite{Kang_ICC15} aim to motivate WiFi Access Point (AP) owners to provide data offloading services for network operators. They proposed an incentive mechanism, where network operators award WiFi AP owners based on both the amount of data offloading services and service quality they provide. This work also focuses on maximizing network operators' utility.

Iosifidis et al. \cite{Iosifidis_CM14} discussed the challenges of designing incentive schemes to encourage WiFi AP owners to participate in offering data offloading services (such networks are called user-provided networks in \cite{Iosifidis_CM14}). They pointed out that energy consumption and cost for providing and receiving offloading services are critical to incentive scheme designs.

However, all above works \cite{Gao_TMC14,Cai_ICC15,Kang_ICC15} focused on maximizing the utility of network operators rather than users and they did not consider the issue of energy consumption of mobile devices, which is a major concern to all battery-powered mobile devices. Different from \cite{Gao_TMC14,Cai_ICC15,Kang_ICC15}, our goal in this paper is to maximize rewards of users under energy consumption constraint. 

There are also some works considering energy consumption issues in offloading cellular traffic. Ra et al. \cite{Ra_MobiSys10} studied the tradeoff between energy consumption and data delay in WiFi offloading networks. They proposed an algorithm that can achieve the minimum energy consumption under queue stability constraint. However, they did not consider incentive issue and their objective is different from ours.

Al-Kanj et al. \cite{Lina_Globecom11} investigated how to offload cellular traffic under energy consumption constraint on mobile devices in multihop ad hoc networks. Their objective is to minimize the usage of scarce cellular spectrum, which is different from the goal of our work.

Recently, Onoue et al. \cite{Onoue_VTC14} considered reducing data delay in WiFi offloading networks under energy consumption constraint. The idea is that a user prefetches the data to be used before he leaves WiFi network coverage. Therefore, the user can obtain the data immediately when he leaves WiFi coverage without waiting for next WiFi connection, thus reducing data delay. However, they did not consider incentive issues.

\section{Network Models}  \label{section:sm}
In this section, we introduce network models for WiFi offloading system concerning network time, packet arrival process, wireless links, rewards and energy consumption.
Key notations are summarized in Table \ref{table:notation}.
\begin{table*}[!ht]
\caption{Key notations in WiFi offloading system.}
\label{table:notation}
\centering
\begin{tabular}{ |l| l| }
\hline
\textbf{Notation} & \textbf{Description} \\ \hline
$t$ & slotted network time, $t=0, 1, 2, \cdots$ \\ \hline
$a(t)$ & the number of packets arriving at the mobile device at time slot $t$ \\ \hline
$Q(t)$ & the number of packets in the queue at the beginning of time slot $t$, also called queue size \\ \hline
$Z(t)$ & the value of virtual queue at the beginning of time slot $t$ \\ \hline
$n$ & the maximum number of wireless links available for the mobile device every time slot $t$ \\ \hline
$\mathbf{S}(t)$ & the vector of states of $n$ wireless links at slot $t$, where element $S_1(t)$ is the state of cellular wireless link \\
                & and all other vector elements are the states of WiFi wireless links\\ \hline
$\boldsymbol{\alpha}(t)$ & a transmission decision made by the mobile device at slot $t$ \\ \hline 
$\mathcal{D}$ & the set of all $n+1$ possible transmission decisions  \\ \hline
$\pi$ & a transmission policy \\ \hline
$\mathcal{P}$ & the set of all possible transmission policies for the considered WiFi offloading network \\ \hline
$\boldsymbol{\alpha}^{\pi}(t)$ & the transmission decision made by policy $\pi$ at slot $t$  \\ \hline
$p^{\pi}(t)$ & the energy consumption under transmission policy $\pi$ at slot $t$ \\ \hline
$p_c$ & the energy consumption when transmitting packets by a cellular wireless link at a time slot  \\ \hline
$p_w$ & the energy consumption when transmitting packets by a WiFi wireless link  at a time slot  \\ \hline
$p^{av}$ & constraint on average energy consumption  \\ \hline
$b^{\pi}(t)$ & the transmission capacity under transmission policy $\pi$ at time slot $t$  \\ \hline
$r^{\pi}(t)$ & the reward given to the user under transmission policy $\pi$ at slot $t$ \\ \hline
\end{tabular}
\end{table*}


\textbf{Network time:} In a WiFi offloading network as illustrated in Figure \ref{fig:WiFioffload}, network time is divided into time slots $t$ of equal duration, $t=0, 1, 2, \cdots$. For example, one time slot could be duration of $20$ seconds.

\textbf{Packet arrival process and queue:} Packets arrive at the mobile device following a general stochastic process $\{a(t); t\ge 0\}$, where $a(t)$ is the number of packets arriving at the mobile device at time slot $t$. We assume $a(t)$ has finite second moment for any $t$, $\mathbb{E}\{a(t)^2\} < \infty$. All arriving packets will be stored in a queue of the mobile device awaiting for wireless transmission. Denote by $Q(t)$ the number of packets in the queue at the beginning of time slot $t$.

\textbf{Wireless links:} If a mobile device is connected through a wireless channel to either a cellular base station (BS) or a WiFi access point (AP), we say a wireless link is available for the mobile device. We assume at most $n>0$ wireless links are available for the mobile device every time slot $t$, including $1$ cellular wireless link and $n-1$ WiFi wireless links. The state of a wireless link at slot $t$ (i.e., how many packets can be transmitted by that link at slot $t$) is a random variable and varies from time slot to time slot, due to factors like user mobility, interference and wireless channel fading. Denote the states of $n$ wireless links at slot $t$ by a vector $\mathbf{S}(t)=\big(S_1(t), S_2(t), \cdots, S_n(t) \big)$, where vector element $S_1(t)$ is the state of the cellular wireless link (i.e., at most $S_1(t)$ packets can be transmitted by cellular link at slot $t$) and $S_i(t), 2\le i \le n,$ is the state of WiFi wireless link $i$ (at most $S_i(t)$ packets can be transmitted by WiFi link $i$ at slot $t$). 

\textbf{Transmission decision:} At each time slot $t$, the mobile device makes a transmission decision from the following $n+1$ decision options:

\textit{D1:} delaying packet transmission (i.e., not transmitting packets at slot $t$),

\textit{D2:} transmitting packets by the cellular link,

\textit{D3:} transmitting packets by a WiFi link $i, 2 \le i \le n$.

We assume the mobile device can transmit packets by at most one wireless link at a time slot. Denote by $\mathcal{D}$ the set of all $n+1$ possible transmission decisions.
Denote a transmission decision at slot $t$ by vector $\boldsymbol{\alpha}(t)=\big(\alpha_1(t), \alpha_2(t), \cdots, \alpha_n(t) \big)$, where $\boldsymbol{\alpha}(t) \in \mathcal{D}$ and $\alpha_i(t)$ is a binary variable regarding wireless link $i$. Then 
\begin{align}
\boldsymbol{\alpha}(t) \!=\!
					\left\{
							\begin{array}{l l}
							  (0, 0, \cdots, 0)& \quad \text{for \textit{D1}}, \\
								(1, 0, \cdots, 0)& \quad \text{for \textit{D2}}, \\
								(0, 0, \cdots, 1,\cdots, 0)& \quad \text{for \textit{D3}, $\alpha_i(t)=1$}.\\
					    \end{array} 
					\right.
\end{align}

\textbf{Transmission policy:} 
A transmission policy $\pi$ is a set of rules for a user to make transmission decisions at each time slot $t$.

%
%
%

For example, a transmission policy $\pi$ can be one that first observes the states of all wireless links $\mathbf{S}(t)$ at each time slot $t$ and then makes a random transmission decision from decision set $\mathcal{D}$. 

Denote by $\boldsymbol{\alpha}^{\pi}(t)=\big(\alpha_1^{\pi}(t), \alpha_2^{\pi}(t), \cdots, \alpha_n^{\pi}(t) \big)$ the transmission decision under transmission policy $\pi$ at time slot $t$, $\boldsymbol{\alpha}^{\pi}(t) \in \mathcal{D}$. Denote by $\mathcal{P}$ the set of all possible policies for the considered WiFi offloading network.

 

\textbf{Energy consumption:} Denote by $p^{\pi}(t)$ the energy consumption under transmission policy $\pi$ at slot $t$.
\begin{align}
p^{\pi}(t) \!=\!
					\left\{
							\begin{array}{l l}
							  0& \quad \text{for \textit{D1}}, \\
								p_c& \quad \text{for \textit{D2}}, \\
								p_w& \quad \text{for \textit{D3}}.\\
					    \end{array} 
					\right.
\end{align}


\textbf{Transmission capacity:} Denote by $b^{\pi}(t)$ the transmission capacity under transmission policy $\pi$ at time slot $t$, i.e., the maximum number of packets that can be transmitted with transmission decision $\boldsymbol{\alpha}^{\pi}(t)$ and link states $\mathbf{S}(t)$, 
\begin{align}
b^{\pi}(t)=\boldsymbol{\alpha}^{\pi}(t)\mathbf{S}(t)=\sum_{i=1}^{n}S_i(t)\alpha_i^{\pi}(t)
\end{align}
We assume $b^{\pi}(t)$ has finite second moment for any $t$ under any transmission policy $\pi \in \mathcal{P}$, i.e., $\mathbb{E}\{b^{\pi}(t)^2\} < \infty$.

\textbf{Incentive:} To encourage mobile device user to offload cellular traffic, one unit of reward $r^{\pi}(t)$ is given to the user if the user decides to delay packet transmission or transmit packets by a WiFi link under transmission policy $\pi$ at slot $t$, i.e., 
\begin{align} \label{eq:rewardDefinition}
r^{\pi}(t) \!=\!
					\left\{
							\begin{array}{l l}
								1& \quad \text{for \textit{D1 or D3}}, \\
								0& \quad \text{for \textit{D2}}.\\
					    \end{array} 
					\right.
\end{align}


\section{Optimal Scheduling Problem Formulation} \label{section:pf}
Our goal is to  maximize user rewards under energy consumption constraint (defined later) in incentive WiFi offloading network.
In this section, we formulate the problem as a stochastic optimization problem \cite{Neely_book10}. 

We first define how queue $Q(t)$ evolves, when queue $Q(t)$ is said to be stable and quantities concerning time average expectations of packet arrivals, energy consumption, transmission capacity and rewards.

\textbf{Queue dynamics:}  Under transmission policy $\pi \in \mathcal{P}$, the queue in the mobile device evolves as follows:
\begin{align}
Q(t+1) &= \max\big[Q(t)-b^{\pi}(t), 0\big] + a(t) \label{eq:qd}
\end{align}

\textbf{Queue stability:} The queueing process $Q(t)$ is mean rate stable \cite{Neely_book10} if 
\begin{align}
\lim_{t \rightarrow \infty} \frac{\mathbb{E}\big\{|Q(t)|\big\}}{t}=0
\end{align}

\textbf{Time average expectations:}
For transmission policy $\pi \in \mathcal{P}$, we define the following time average expectations:
\begin{align}
\overline{a}(t)&=\frac{1}{t}\sum_{\tau=0}^{t-1}\mathbb{E}\{a(\tau)\} \label{eq:avg-packet}\\
\overline{p}^{\pi}(t)&=\frac{1}{t}\sum_{\tau=0}^{t-1}\mathbb{E}\{p^{\pi}(\tau)\} \label{eq:avg-power}\\
\overline{b}^{\pi}(t)&=\frac{1}{t}\sum_{\tau=0}^{t-1}\mathbb{E}\{b^{\pi}(\tau)\} \label{eq:avg-tansCap}\\
\overline{r}^{\pi}(t)&=\frac{1}{t}\sum_{\tau=0}^{t-1}\mathbb{E}\{r^{\pi}(\tau)\} \label{eq:avg-reward}
\end{align}
Note that $0 \le \overline{r}^{\pi}(t) \le 1$ from definition (\ref{eq:rewardDefinition}).

\textbf{Optimal scheduling problem formulation:}
In the considered incentive WiFi offloading network, our objective is to propose a scheduling policy to maximize the time average expected rewards while meeting energy consumption constraint $p^{av}, 0<p^{av}<\infty$, (\ref{P1:constraint1}) and queue stability constraint (\ref{P1:constraint2}). We formulate this problem as a stochastic optimization problem \cite{Neely_book10}:
\begin{align}
& \underset{\pi \in \mathcal{P}}{\text{maximize}} & & \limsup_{t \rightarrow \infty} \overline{r}^{\pi}(t) \\
& \text{subject to} & & \text{1)} \; \limsup_{t \rightarrow \infty} \overline{p}^{\pi}(t) \le p^{av} \label{P1:constraint1}\\
& & & \text{2)} \; \text{The queueing process $Q(t)$ is} \nonumber \\
& & & \quad \; \text{mean rate stable} \label{P1:constraint2}\\
& & & \text{3)} \; \boldsymbol{\alpha}^{\pi}(t) \in \mathcal{D},  \; \forall \; t\ge 0 \;  \label{P1:constraint3}
\end{align}
Note that the maximum value of $\limsup_{t \rightarrow \infty} \overline{r}^{\pi}(t)$ is $1$.

If we let 
\begin{align}
f^{\pi}(t)&=-r^{\pi}(t) \\
y^{\pi}(t)&=p^{\pi}(t)-p^{av}  \label{eq:ypi}
\end{align}
and similarly define their time average expectations $\overline{f}^{\pi}(t), \overline{y}^{\pi}(t)$ like (\ref{eq:avg-packet}) (\ref{eq:avg-power}) (\ref{eq:avg-tansCap}) (\ref{eq:avg-reward}), then the above stochastic optimization problem is equivalent to the following one.
\begin{align}
& \underset{\pi \in \mathcal{P}}{\text{minimize}} & & \liminf_{t \rightarrow \infty} \overline{f}^{\pi}(t) \label{P2:of}\\
& \text{subject to} & & \text{1)} \; \limsup_{t \rightarrow \infty} \overline{y}^{\pi}(t) \le 0 \label{P2:constraint1}\\
& & & \text{2)} \; \text{The queueing process $Q(t)$ is} \nonumber \\
& & & \quad \; \text{mean rate stable} \label{P2:constraint2}\\
& & & \text{3)} \; \boldsymbol{\alpha}^{\pi}(t) \in \mathcal{D},  \; \forall \; t\ge 0 \;  \label{P2:constraint3}
\end{align}

\textbf{Feasible problem:} The above stochastic optimization problem is said to be feasible if there exists at least one policy $\pi \in \mathcal{P} $ that meets all constraints (\ref{P2:constraint1})-(\ref{P2:constraint3}). 
Throughout this paper we only consider feasible problem.

\section{Optimal Scheduling Policy under Energy Constraint} \label{section:OPEC}
In this section, we propose a scheduling policy that meets constraints (\ref{P2:constraint1})-(\ref{P2:constraint3}) and achieves the minimum limit value of $\overline{f}^{\pi}(t)$ (i.e, maximum limit value of rewards $\overline{r}^{\pi}(t)$). Our policy is named Optimal scheduling Policy under Energy Constraint (OPEC), OPEC $\in \mathcal{P}$.


We need several definitions in describing OPEC. First, define a virtual queueing process $\{Z(t); t \ge 0\}$, which is used in OPEC to meet the average energy consumption constraint (\ref{P2:constraint1}). The virtual queueing process evolves as follows:
\begin{align}
Z(t+1)=\max \big[Z(t)+y^{OPEC}(t), 0\big], \label{eq:vq}
\end{align}
where $y^{OPEC}(t)$ is defined in $(\ref{eq:ypi})$ and we assume the initial conditions $Z(0)\ge 0$ and $\mathbb{E}\{Z(0)^2\}<\infty$ \footnote{These two assumptions about $Z(0)$ will be used in proving the optimality of OPEC. See \cite{Neely_book10} for more details.}. 

Then, define the following auxiliary quantities for transmission decision option $\boldsymbol{\alpha}(t)\in \mathcal{D}$, which will help OPEC in making decisions.
\begin{align}
p(t) &=
					\left\{
							\begin{array}{l l}
							  0& \quad \text{for \textit{D1}}, \\
								p_c& \quad \text{for \textit{D2}}, \\
								p_w& \quad \text{for \textit{D3}}.\\
					    \end{array} 
					\right. \\
b(t)&=\boldsymbol{\alpha}(t)\mathbf{S}(t)=\sum_{i=1}^{n}S_i(t)\alpha_i(t) \\
r(t) &=
					\left\{
							\begin{array}{l l}
								1& \quad \text{for \textit{D1 or D3}}, \\
								0& \quad \text{for \textit{D2}}.\\
					    \end{array} 
					\right.
\end{align}

The full description of OPEC policy is given in Algorithm \ref{algorithm:OPEC}.

\textbf{Basic idea of OPEC:} To meet queue stability constraint (\ref{P2:constraint2}), it is sufficient for one policy to greedily minimize $-Q(t)b(t)$ every time slot $t$ as proved in \cite{Neely_book10} ($b(t), f(t), y(t)$ are defined in Algorithm \ref{algorithm:OPEC}). Similarly, to meet the constraint on average energy consumption (\ref{P2:constraint1}), it suffices to greedily minimize $Z(t)y(t)$ every time slot $t$ \cite{Neely_book10}. Thus, to meet both constraints (\ref{P2:constraint2}) and (\ref{P2:constraint1}), OPEC greedily minimizes $-Q(t)b(t)+Z(t)y(t)$ every time slot $t$. 
Since our final goal is to minimize objective function (\ref{P2:of}) while meeting constraints (\ref{P2:constraint1})-(\ref{P2:constraint3}), OPEC also needs to consider $f(t)$. To account for $f(t)$, OPEC minimizes a weighted sum $V f(t)-Q(t)b(t)+Z(t)y(t)$ every time slot $t$. 

Control parameter $V$ measures the importance of getting rewards: larger $V$ means we prefer getting more rewards, which comes at the cost of an increase in queue size $Q(t)$. Thus, by controlling $V$ we can achieve a tradeoff between queue size and rewards.
It can be proved, by Lyapunov optimization theory \cite{Neely_book10}, that OPEC policy indeed achieves the minimum limit value of $\overline{f}^{\pi}(t)$ while meeting constraints (\ref{P2:constraint1})-(\ref{P2:constraint3}) by greedily minimizing the above weighted sum.

\begin{algorithm}[!ht]
\caption{OPEC scheduling policy}
\label{algorithm:OPEC}
\begin{algorithmic}[1]
\REQUIRE  $Q(t):$ the number of packets in the queue at slot $t$ \\
         \quad  \; $Z(t):$ the value of the virtual queue at slot $t$ \\
         \quad  \; $\mathbf{S}(t):$ wireless link states at slot $t$ \\
         \quad  \; $a(t):$ the number of packets arriving at $t$\\
\renewcommand{\algorithmicrequire}{\textbf{Parameters:}} 
\REQUIRE $n:$ the maximum number of available links \\
         \quad  \; $p_w:$ energy consumption of transmitting packets \\
         \quad \quad\quad\; \;  by a WiFi link\\
         \quad  \; $p_c:$ energy consumption of transmitting packets \\
         \quad \quad\quad\; \;  by a cellular link \\
         \quad  \; $V:$ control parameter $(V \ge 0)$\\
\ENSURE A transmission decision $\boldsymbol{\alpha}^{OPEC}(t) \in \mathcal{D}$ 

\STATE $value = \infty$;
\STATE $\boldsymbol{\alpha}^{OPEC}(t)=(0,0, \cdots,0)$;
\STATE $tmp=0$;
	\FOR{\textbf{all} \text{transmission decision options $\boldsymbol{\alpha}(t)\in \mathcal{D}$}}
	    \STATE $f(t)=-r(t)$;
			\STATE $b(t)=\sum_{i=1}^{n}S_i(t)\alpha_i(t)$;
			\STATE $y(t)=p(t)-p^{av}$;
			\STATE Calculate $tmp = V f(t)-Q(t)b(t)+Z(t)y(t)$; \\
			\IF {$tmp<value$}
					\STATE $value=tmp$ \, ;
					\STATE $\boldsymbol{\alpha}^{OPEC}(t)=\boldsymbol{\alpha}(t)$;
			\ENDIF
	\ENDFOR
\STATE $Q(t+1) = \max\big[Q(t)-b^{OPEC}(t), 0\big] + a(t)$;
\STATE $Z(t+1)=\max \big[Z(t)+y^{OPEC}(t), 0\big]$;
\RETURN $\boldsymbol{\alpha}^{OPEC}(t)$.
\end{algorithmic}
\end{algorithm}

\textbf{Line 1-3:} Initialize the decision $\boldsymbol{\alpha}^{OPEC}(t)$ to be not transmitting packets. $value$ and $tmp$ are auxiliary variables to be used in the following steps.

\textbf{Line 4-13:} For each time slot $t$, OPEC scheduling policy looks for a transmission decision $\boldsymbol{\alpha}(t)\in \mathcal{D}$ that minimizes 
\begin{align}
tmp=V f(t)-Q(t)b(t)+Z(t)y(t). \label{eq:value}
\end{align}

\textbf{Line 14:} Update the value of queue $Q(t)$.

\textbf{Line 15:} Update the value of virtual queue $Z(t)$.

\section{Simulation validation} \label{section:sv}
In this section, we verify our OPEC scheduling policy by simulations.

\subsection{Simulation settings}
For simulating the considered WiFi offloading network, we developed a customized simulator, which takes as inputs packet arrival process $\{a(t); t \ge 0\}$, maximum number of wireless links $n$, stochastic process of wireless link states $\mathbf{S}(t)$, energy consumption $p_c, p_w$, energy consumption constraint $p^{av}$ and control parameter $V$. By configuring these parameters, we can simulate a wide range of network scenarios.

For illustration purpose, we consider a network scenario with the following network settings.
\begin{itemize}
	\item The number $a(t)$ of packets arriving at the queue for any time slot $t\ge 0$ is an Independent and Identically Distributed (IID) random variable, where $Pr\{a(t)=0\}=0.2,  Pr\{a(t)=2\}=0.3,Pr\{a(t)=3\}=0.5$.
	\item There are at most two wireless links ($n=2$) available every time slot: one cellular link and one WiFi link.
	\item Wireless link state vector $\mathbf{S}(t)$ is an IID vector random variable over time slots. Since cellular network is of low data rates and almost always available, we set cellular link states and probabilities as $Pr\{S_{1}(t)=0\}=0.1, Pr\{S_{1}(t)=1\}=0.2, Pr\{S_{1}(t)=2\}=0.7$. Since WiFi network is of high data rates and intermittently available, we set WiFi link states and probabilities as $Pr\{S_{2}(t)=0\}=0.7, Pr\{S_{2}(t)=2\}=0.05, Pr\{S_{2}(t)=4\}=0.05, Pr\{S_{2}(t)=10\}=0.1, Pr\{S_{2}(t)=20\}=0.1$.
	\item For energy consumption, we set $p_c=1.15$ J/slot when transmitting by cellular link and $p_c=1.1$ J/slot when transmitting by WiFi link. Constraint on average energy consumption is set to be $p^{av}=0.8$ J/slot.

\end{itemize}
We vary control parameter $V$ in our simulations and examine its impact on performance metrics like time average queue size $\overline{Q}$, time average rewards $\overline{r}^{OPEC}$ and time average energy consumption $\overline{p}^{OPEC}$ defined below
\begin{align}
\overline{Q}&=\frac{1}{t}\sum_{\tau=0}^{t-1}Q(\tau) \\
\overline{r}^{OPEC}&=\frac{1}{t}\sum_{\tau=0}^{t-1}r^{OPEC}(\tau) \\
\overline{p}^{OPEC}&=\frac{1}{t}\sum_{\tau=0}^{t-1}p^{OPEC}(\tau)
\end{align}

\subsection{Simulation results}
\begin{figure}[!th]
\centering
\includegraphics[width=4in]{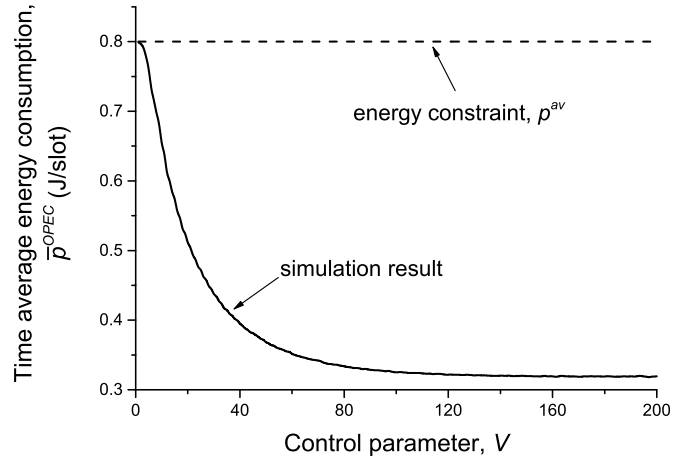}
\caption{Time average energy consumption $\overline{p}^{OPEC}$ vs. control parameter $V$.}
\label{fig:energyV}
\end{figure}

\begin{figure}[!th]
\centering
\includegraphics[width=4in]{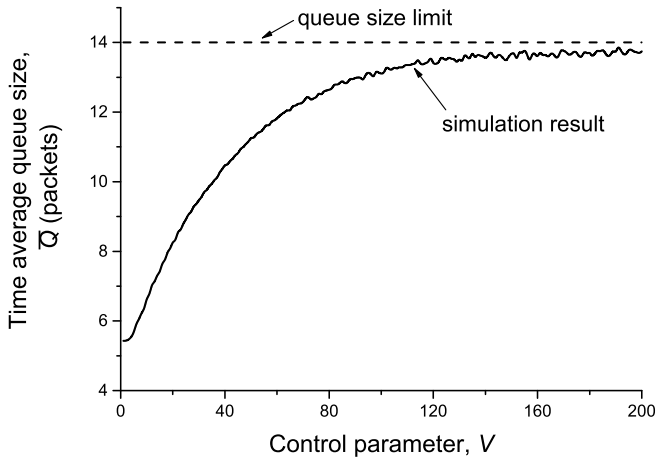}
\caption{Time average queue size $\overline{Q}$ vs. control parameter $V$.}
\label{fig:queueV}
\end{figure}

\begin{figure}[!th]
\centering
\includegraphics[width=4in]{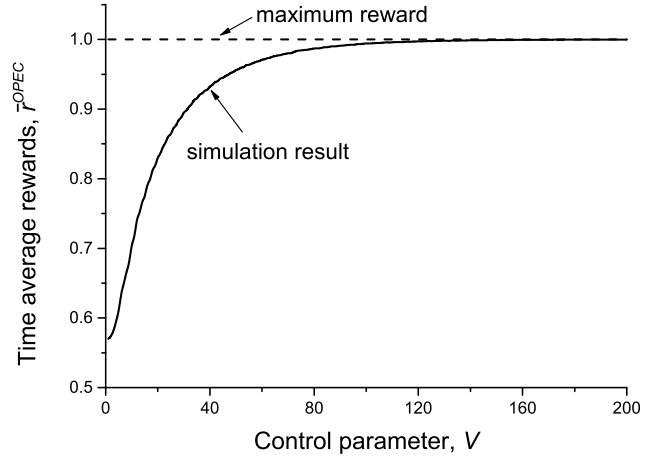}
\caption{Time average rewards $\overline{r}^{OPEC}$ vs. control parameter $V$.}
\label{fig:rewardV}
\end{figure}

After running simulations for $10^6$ slots (i.e., $11.5$ days if $1$ slot is equal to $1$ second) under different settings of $V$, we collected data of $Q(t), r^{OPEC}(t), p^{OPEC}(t)$ and calculated their time averages $\overline{Q}, \overline{r}^{OPEC}, \overline{p}^{OPEC}$, as summarized in Figures \ref{fig:energyV}, \ref{fig:queueV}, \ref{fig:rewardV}.

We first check whether OPEC satisfies the constraint on average energy consumption. Figure \ref{fig:energyV} shows the simulated results on $\overline{p}^{OPEC}$ as $V$ increases from $1$ to $200$. From Figure \ref{fig:energyV}, we can see that the time average energy consumption $\overline{p}^{OPEC}$ under OPEC scheduling policy is always less than the constraint $p^{av}=0.8$ J/slot. Thus, OPEC meets the given constraint on average energy consumption. Further observation of Figure \ref{fig:energyV} shows that as $V$ increases, $\overline{p}^{OPEC}$ decreases and approaches a limit value $0.32$, indicating that with constraints on queue stability and average energy consumption, there exists a minimum value $p^{min}$ of average energy consumption for the considered WiFi offloading network.
That is to say, the average energy consumption constraint $p^{av}$ should be larger than $p^{min}$, otherwise no scheduling policy can meet average energy consumption constraint while keeping queue stable.

We then check whether OPEC satisfies the queue stability constraint. Figure \ref{fig:queueV} shows the simulated results on $\overline{Q}$ as $V$ increases. From Figure \ref{fig:queueV}, we can see that the time average queue size $\overline{Q}$ under OPEC scheduling policy increases as $V$ increases and tends to reach a limit value that is less than $14$. As a common practice, we estimate $\mathbb{E}\big\{|Q(t)|\big\}$ by time average $\overline{Q}$. Therefore,
\begin{align}
\lim_{t \rightarrow \infty} \frac{\mathbb{E}\big\{|Q(t)|\big\}}{t}&=\lim_{t \rightarrow \infty} \frac{\overline{Q}}{t} \le \lim_{t \rightarrow \infty} \frac{14}{t}=0 
\end{align}
Thus, OPEC also meets the constraint of queue being mean rate stable.

Finally, we check whether OPEC can achieve the maximum average reward $1$. Figure \ref{fig:rewardV} shows simulated results on $\overline{r}^{OPEC}$ as $V$ increases. We can see from the figure that average reward $\overline{r}^{OPEC}$ under OPEC scheduling policy increases as $V$ increases and reaches the optimal value 1, thus verifying that OPEC can indeed achieve the maximum average reward $1$. In summary, by simulations we validated that OPEC scheduling policy can achieve the maximum average rewards while satisfying constraints on queue stability and energy consumption.


\section{Conclusion} \label{section:conclusion}
In this paper, we focused on maximizing users' rewards in incentive WiFi offloading under energy consumption constraint. We proposed a flexible scheduling policy, called Optimal scheduling Policy under Energy Constraint (OPEC). 
As validated by simulation, OPEC can achieve the maximum rewards while keeping queue stable and meeting energy consumption constraint.
One merit of OPEC is that it automatically adapts to random packet arrivals, intermittent WiFi connection and time varying wireless link states, not requiring a priori knowledge of packet arrival and wireless link probabilities. 
As a future work, we will consider offloading cellular traffic by multiple complementary networks, such as WiFi, WiMax and Femtocell networks.

\bibliographystyle{IEEEtran}
\bibliography{reference}

\end{document}